\documentclass[footinbib,twocolumn,english,aps,prl,superscriptaddress]{revtex4-1}
\usepackage[T1]{fontenc}
\usepackage[latin9]{inputenc}
\usepackage{endnotes}
\usepackage{amsmath}
\usepackage{amssymb}
\usepackage{graphicx}

\makeatletter
\let\footnote=\endnote

\renewcommand{\fnum@figure}{FIG.~\thefigure}
\usepackage{hyperref}
\hypersetup{colorlinks=true,linkcolor=blue,citecolor=blue,urlcolor=blue}
\makeatother

\usepackage{babel}
\begin{document}
\title{Electronic nematicity in URu$_{2}$S\textmd{\normalsize{}i$_{2}$}
revisited}
\author{Liran Wang}
\thanks{These authors contributed equally to this work.}
\affiliation{\textsuperscript{}Institute for Quantum Materials and Technologies, Karlsruhe Institute
of Technology, 76021 Karlsruhe, Germany}
\author{Mingquan He}
\thanks{These authors contributed equally to this work.}
\affiliation{\textsuperscript{}Institute for Quantum Materials and Technologies, Karlsruhe Institute
of Technology, 76021 Karlsruhe, Germany}
\author{Fr$\acute{\rm e}$d$\acute{\rm e}$ric Hardy}
\affiliation{\textsuperscript{}Institute for Quantum Materials and Technologies, Karlsruhe Institute
of Technology, 76021 Karlsruhe, Germany}
\author{Dai Aoki}
\affiliation{\textsuperscript{}Universit$\acute{\rm e}$ Grenoble Alpes, CEA, PHELIQS, 38000
Grenoble, France}
\affiliation{Institute for Materials Research, Tohoku University, Oarai, Ibaraki
311-1313, Japan}
\author{Kristin Willa}
\affiliation{\textsuperscript{}Institute for Quantum Materials and Technologies, Karlsruhe Institute
of Technology, 76021 Karlsruhe, Germany}
\author{Jacques Flouquet}
\affiliation{\textsuperscript{}Universit$\acute{\rm e}$ Grenoble Alpes, CEA, PHELIQS, 38000
Grenoble, France}
\author{Christoph Meingast}
\email{christoph.meingast@kit.edu}

\affiliation{\textsuperscript{}Institute for Quantum Materials and Technologies, Karlsruhe Institute
of Technology, 76021 Karlsruhe, Germany}
\date{23/4/20}
\begin{abstract}
The nature of the  hidden-order (HO) state in
URu$_{2}$S{\normalsize{}i$_{2}$} remains one of the major unsolved
issues in heavy-fermion physics. Recently, torque magnetometry, x-ray
diffraction and elastoresistivity data have suggested that the HO
phase transition at $T_{HO}\thickapprox$ 17.5 K is driven by electronic nematic effects. Here, we search for thermodynamic signatures
of this purported structural instability using anisotropic thermal-expansion,
Young\textquoteright s modulus, elastoresistivity and specific-heat
measurements. In contrast to the published results, we find no evidence
of a rotational symmetry-breaking in any of our data. Interestingly,
our elastoresistivity measurements, which are in full agreement with
published results, exhibit a Curie-Weiss divergence, which we however
attribute to a volume and not to a symmetry-breaking effect. Finally,
clear evidence for thermal fluctuations is observed
in our heat-capacity data, from which we estimate the HO correlation
length. 
\end{abstract}
\maketitle
Despite 35 years of intensive experimental and theoretical efforts
\cite{Mydosh11_review_URu2Si2,Mydosh14_Review_URu2Si2}, the microscopic
nature of the hidden-order state (HO) in URu$_{2}$Si$_{2}$ is unknown
and  remains
one of the major unsolved issues in heavy-fermion physics. Recently,
torque magnetometry, x-ray diffraction and elastoresistivity, have
reported experimental signatures of electronic nematicity
at the hidden-order phase transition $T_{HO}\thickapprox$ 17.5 K,
which, when confirmed, would narrow down the possible order parameters \cite{Okazaki11_Torque_URu2Si2,Tonegawa14_Nematicity_URu2Si2,Riggs15_Nematicity_URu2Si2}.
These reports point to a crystallographic symmetry
lowering at the HO transition from a tetragonal to an orthorhombic
structure, which would favor several theories, e.g. those involving
multipolar orders, which rely on the breaking of the 4-fold symmetry
below $T_{HO}$ (see Ref. \cite{Ikeda2012}). However, another recent x-ray diffraction study found no evidence of a structural transition \cite{2014tabata}, and improved nuclear-magnetic resonance (NMR) experiments now suggested an odd-parity electronic multipolar ordering within a tetragonal environment \cite{KambePRB97NMR}, although previous NMR data pointed to a two-fold ordering \cite{KambePRB2015}.

In this Letter, in order to resolve the above controversy, we use three different sensitive experimental techniques to search for bulk experimental evidence of
the purported nematic order parameter in well-characterized URu$_{2}$Si$_{2}$ single crystals \cite{BourdarotJPSJ2010}. First, we utilize anisotropic high-resolution capacitance dilatometry, which is several orders of magnitude more sensitive than the x-ray
diffraction measurements reported in Ref. \cite{Tonegawa14_Nematicity_URu2Si2},
and has recently been used to study nematicity
 Fe-based systems \cite{Anna_FeSe_PRL2015,Wang18_Nematicity_C4_Phase,Liran_JPSJ2019}. Second, a symmetry-breaking
transition inevitably leads to a drastic softening of its associated
shear modulus, and we have thus performed Young\textquoteright s
modulus measurements using a three-point-bending setup, which has
been shown to be a very sensitive technique for detecting lattice
softening in Fe-based materials \cite{Boehmer14_Nematic_Susceptibility,Anna_FeSe_PRL2015,Boehmer16_Nematicity_CR,Wang18_Nematicity_C4_Phase}.
Finally, we study the elastoresistivity \cite{ChuScience2012} as a third sensitive method
for observing nematicity. Our
main result is that we find absolutely no evidence for a symmetry-breaking
transition in either the thermal expansion or the Young's modulus
measurements. Our elastoresistivity data interestingly exhibit a Curie-Weiss divergence,
similar to the results of Riggs \textit{et al.} \cite{Riggs15_Nematicity_URu2Si2}.
However, we find that a very similar behavior can be inferred from
hydrostatic-pressure measurements, revealing that a Curie-Weiss-like
response under an enforced symmetry-breaking strain does not necessarily
imply a nematic origin.

Single crystals of URu$_{2}$Si$_{2}$ were prepared by the Czochralski
method and annealed at high temperature under ultra-high vacuum, as described in details in Refs \cite{Aoki10_Reentrance_U122,MatsudaJPSJ2011}. The residual resistivity ratio (RRR) is typically around 100, indicating high-quality crystals.
Thermal-expansion measurements were carried out on a single crystal
(2.0 mm $\times$ 1.8 mm $\times$ 2.0 mm) using a home-built high-resolution capacitance
dilatometer \cite{Meingast91_Thermal-Epansion_YBCO}. Heat-capacity
measurements were made on the same crystal with the Physical Properties
Measurement System (PPMS) from Quantum design using the dual-slope
method \cite{Marcenat_Thesis,Riegel86_Dual_Slope_SHC}. Young\textquoteright s
modulus data were obtained with the same dilatometer set up in a three-point-bending
configuration \cite{Boehmer14_Nematic_Susceptibility,Boehmer16_Nematicity_CR}.
A sketch of this setup, in which the force from the dilatometer
springs causes a deflection of the crystal, is shown in Fig.\ref{Fig3}. Elastoresistivity
measurements were made by gluing a crystal on a glass-fiber-reinforced-plastic
substrate as described in Ref. \cite{He17_GFK_Ba122} and in the supplemental section \cite{NoteX}

\begin{figure}[t]
\begin{centering}
\includegraphics[scale=0.7]{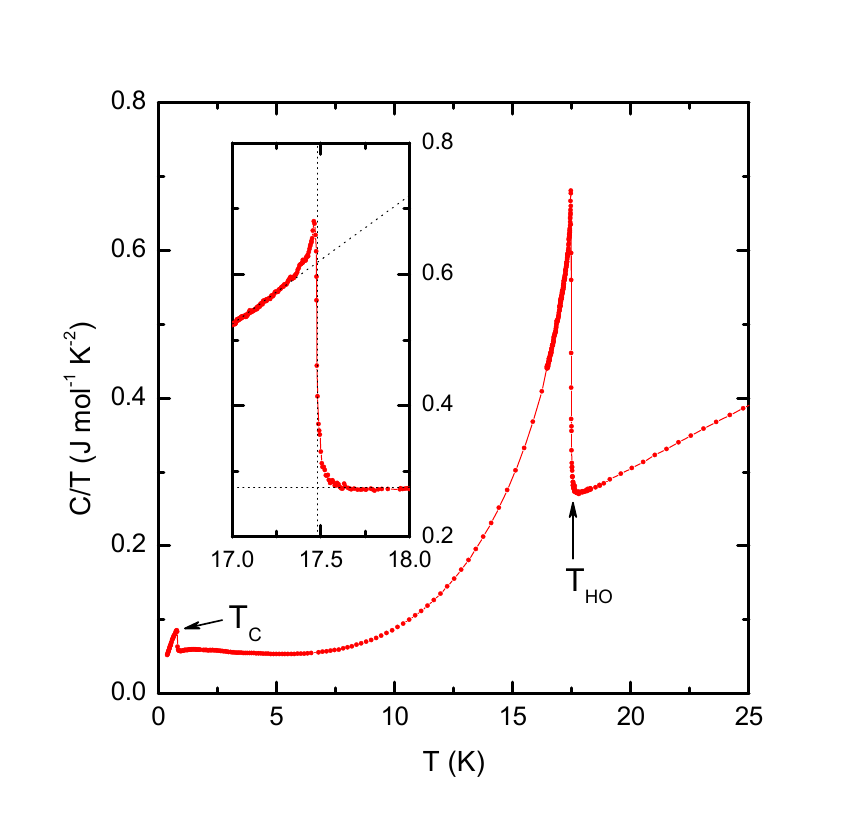}
\par\end{centering}
\caption{Low-temperature heat capacity of our URu$_{2}$Si$_{2}$ single
crystal. The data around $T_{HO}$ = 17.5 K are plotted on an enlarged scale in the inset.
The high quality of our single crystal is attested by the sharpness
of the transition, allowing thermal fluctuations of the HO order parameter to be 
clearly observed. \label{Fig1}}
\end{figure}

The HO transition results in a prominent anomaly at $T_{HO}$ = 17.49
K in the heat capacity as illustrated in Fig. \ref{Fig1}. Clear signs of order-parameter fluctuations in a range of
$\thickapprox$ 0.1 K above and below $T_{HO}$ (see inset of Fig.\ref{Fig1}) are observed, from which we estimate a Ginzburg parameter of G$_{i}$ = 0.006.
To the best of our knowledge, this is the first time that such fluctuations
have been observed, and from G$_{i}$ and the 'condensation energy' H$_{c}$ = 0.47 T (derived from specific heat), we estimate a correlation length of 1 nm using standard Gaussian fluctuation theory.  This short correlation length is an indication of a very short-range interaction leading to the HO state. 
From the rounding of the anomaly, we estimate that the
transition width is as small as 0.015 K, clearly demonstrating the
high homogeneity of our crystals. Systematic studies on different crystals have shown that the HO phase is very robust, i.e. not strongly impurity dependent \cite{MatsudaJPSJ2011}.

\begin{figure}[b]
\includegraphics[scale=.8]{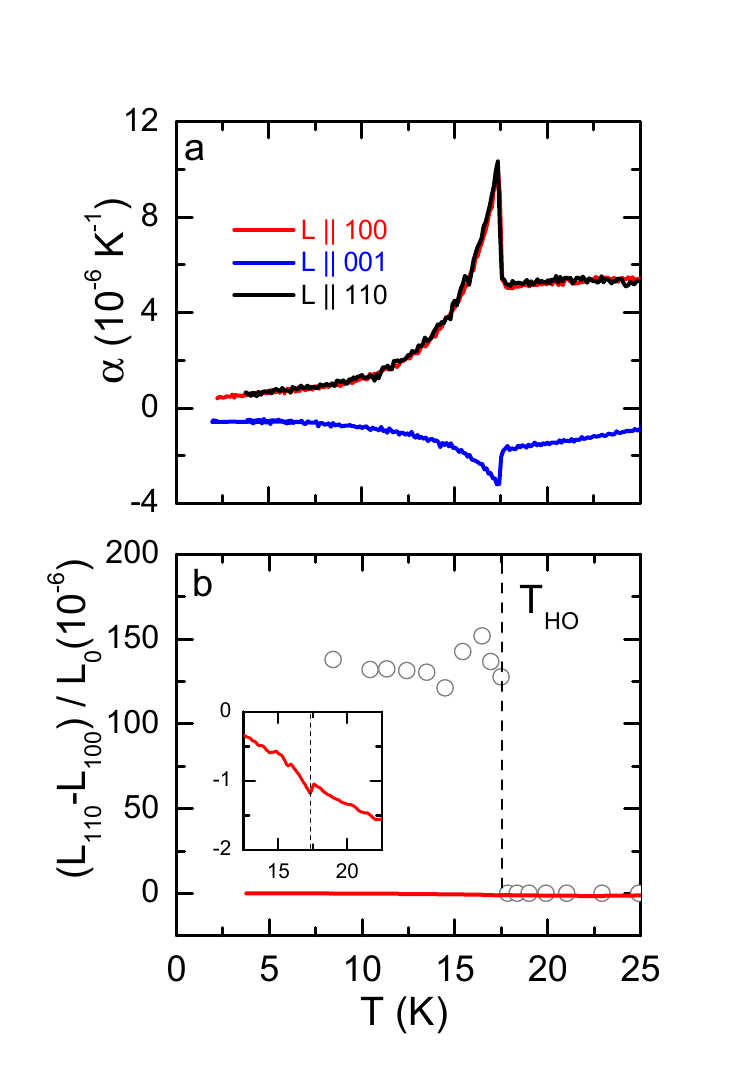}

\caption{(a) Thermal-expansion coefficients along {[}100{]}, {[}001{]} and
{[}110{]} directions near the hidden-order transition. (b) Orthorhombic
distortion, i.e. the difference in thermal expansion along {[}100{]}
and {[}110{]} compared to the x-ray diffraction data of Tonegawa \textit{et
al.} \cite{Tonegawa14_Nematicity_URu2Si2} (open circles). No symmetry-breaking distortion
can be observed in our data (solid red line).\label{Fig2}}
\end{figure}

\begin{figure}
\includegraphics[scale=.30]{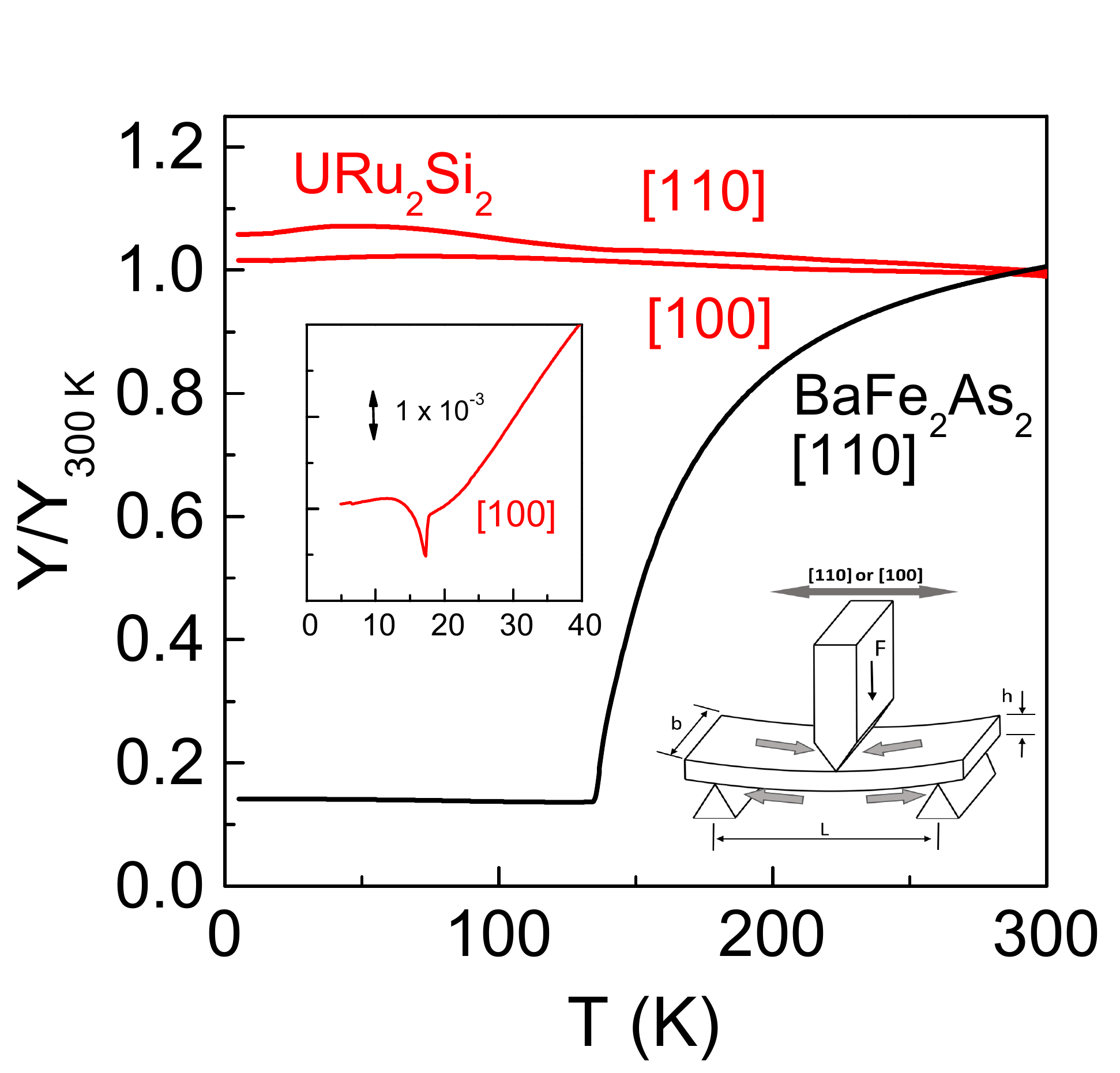}

\caption{Young's moduli of URu$_{2}$Si$_{2}$ measured along the {[}100{]}
and {[}110{]} directions using a three-point bending setup (see right inset). In contrast to BaFe$_{2}$As$_{2}$, we
find absolutely no evidence for any soft-mode in URu$_{2}$Si$_{2}$.
The left inset shows a magnified view of the {[}100{]} data around  $T_{HO}$.\label{Fig3}}
\end{figure}

A large $C_{4}$-symmetry-breaking strain, as reported by x-ray diffraction
\cite{Tonegawa14_Nematicity_URu2Si2} below $T_{HO}$, should be easily
detected using our dilatometer by comparing the strains $\varepsilon_{100}(T)=\left(\frac{\Delta L_{100}}{L_{100}}\right)$
and $\varepsilon_{110}(T)=\left(\frac{\Delta L_{110}}{L_{110}}\right)$
measured respectively along the {[}100{]} and {[}110{]} directions,
as has been demonstrated for Fe-based materials \cite{Boehmer13_FeSe_Growth,Boehmer15_C4phase_BaK122,Wang16_Entropy_C4_phase}.
This is because our spring-loaded dilatometer exerts a non-negligible
stress along the measurement direction, and thus, for a measurement
along the tetragonal {[}110{]} direction, the population of possible
structural domains (twins) with the shorter orthorhombic axis should
be favored by this stress. This would result in an in-situ detwinning
of the sample below $T_{HO}$, if the crystal symmetry were
lowered. On the other hand, the twin population would remain unaffected
by the dilatometer force for measurements along the {[}100{]} direction,
which probe a mixture of both orthorhombic axes. We note that for
the present measurements we have used \textquoteleft hard\textquoteright{}
springs, which apply a large force of about 300 g (3 MPa) to the crystal,
i.e. about a factor of 5 larger than in our previous experiments on
pnictides \cite{Boehmer13_FeSe_Growth,Boehmer15_C4phase_BaK122,Wang16_Entropy_C4_phase}.
As illustrated in Fig. \ref{Fig2}(a), we find no measurable difference
between the coefficients of linear thermal expansion $\alpha_{100}(T)=\left(\frac{\partial\varepsilon_{xx}}{\partial T}\right)$
and $\alpha_{110}(T)=\left(\frac{\partial\varepsilon_{xy}}{\partial T}\right)$ measured along the {[}100{]} and {[}110{]} directions, respectively.
To quantitatively compare our data with the x-ray diffraction results
\cite{Tonegawa14_Nematicity_URu2Si2}, we plot in Fig. \ref{Fig2}(b)
the orthorhombic distortion, i.e. the normalized difference in length
$\frac{L_{110}-L_{100}}{L_{0}}$. Clearly, our data,
which show no signature of any distortion (see red line), are incompatible
with the large distortion reported in Ref. \cite{Tonegawa14_Nematicity_URu2Si2}.
Our resolution limit is about 700 times smaller than the reported
distortion (see inset in Fig. \ref{Fig2}(b)\cite{NoteX}. We note that our results are in excellent agreement with
those of de Visser et al. \cite{deVisser86_Alpha_URu2Si2,Kuwahara97_Alpha_elastic_URu2Si2}
published more than two decades ago. Additionally, the diffraction-inferred
strain changes discontinuously at $T_{HO}$, which seems at odds with
the clear second-order nature of the transition, observed in both
specific-heat and thermal-expansion measurements Figs. \ref{Fig1} and \ref{Fig2}(a).

Another powerful method for searching for a nematic
instability is to measure the relevant shear modulus ($c_{66}$ elastic constant for URu$_{2}$Si$_{2}$), which necessarily
has to approach zero as the transition is approached from above. 
As shown
previously for BaFe$_{2}$As$_{2}$, the appropriate Young's modulus
is in a good approximation proportional to $c_{66}$ \cite{Boehmer14_Nematic_Susceptibility,Anna_FeSe_PRL2015,Boehmer16_Nematicity_CR,Wang18_Nematicity_C4_Phase}.
For URu$_{2}$Si$_{2}$ the relevant Young\textquoteright s moduli,
$Y_{100}$ and $Y_{110}$, can be determined using the three-point-bending
technique with the tetragonal {[}100{]} and {[}110{]} crystal axis
perpendicular to the beam supports, respectively (see inset of Fig. \ref{Fig3}).
Here the moduli are expressed in terms of the elastic constants $c_{ij}$
by

\begin{align*}
Y_{100} & =\left(c_{11}-c_{12}\right)\left(1+\eta\right),
\end{align*}
and 
\begin{eqnarray*}
Y_{110} & = & \left(\frac{1}{c_{66}}+\frac{1}{\gamma}\right)^{-1}
\end{eqnarray*}
with $\eta=\frac{c_{12}c_{33}-c_{13}^{2}}{c_{11}c_{33}-c_{13}^{2}}$
and $\gamma=\frac{c_{11}+c_{12}}{2}-\frac{c_{13}^{2}}{c_{33}}$. Near
a structural transition the shear mode will soften significantly,
 with $c_{66}<<\gamma$ and thus
\begin{eqnarray*}
Y_{110} & \propto & c_{66}
\end{eqnarray*}

In Fig. \ref{Fig3}, we compare $Y_{100}$ and $Y_{110}$ of URu$_{2}$Si$_{2}$
to the soft shear mode of BaFe$_{2}$As$_{2}$ ($Y_{110}$). In contrast to
the strong softening observed in BaFe$_{2}$As$_{2}$ at the spin-density-wave
transition $T_{s,N}$ = 140 K, both Young's moduli of URu$_{2}$Si$_{2}$ increase upon cooling, as expected for phonon hardening, and exhibit
absolutely no evidence for any kind of soft-mode behavior. In Ref\cite{Tonegawa14_Nematicity_URu2Si2} it was argued that the anti-ferro nature of HO makes the elastic constant $C_{11}$ - $C_{12}$,
being sensitive to Q = 0, to only couple weakly to the proposed
symmetry breaking. Even for weak coupling, one would nevertheless expect a drastic softening of the relevant elastic mode, however in a smaller temperature interval. 

We note
that the top (bottom) surface of the bar-shaped sample experiences
even larger compressive (tensile) stresses ($\approx$ 15 MPa) in
these bending configuration than in our thermal-expansion measurements \cite{Wang18_Nematicity_C4_Phase}, and,  because we see no softening even under these
extreme conditions, we conclude that there exists no evidence for
a $C_{4}$-symmetry reduction in URu$_{2}$Si$_{2}$ even under the quite
large uniaxial strains of 3 - 15 MPa. As seen in Fig. \ref{Fig3}, both moduli
do soften very slightly below approximately 70 K, as reported earlier
for $\frac{c_{11}-c_{12}}{2}$  by ultrasound investigations, which however is not indicative of a soft mode, but rather indicates
a coupling of the lattice to the quadrupolar moment as argued by Kuwahara
\textit{et al.} \cite{Kuwahara97_Alpha_elastic_URu2Si2}. Additionally there is the small expected softening at $T_{HO}$ resulting from the pressure or strain dependence of $T_{HO}$ (see inset of Fig. \ref{Fig3}).

\begin{figure*}
\includegraphics[scale=.40]{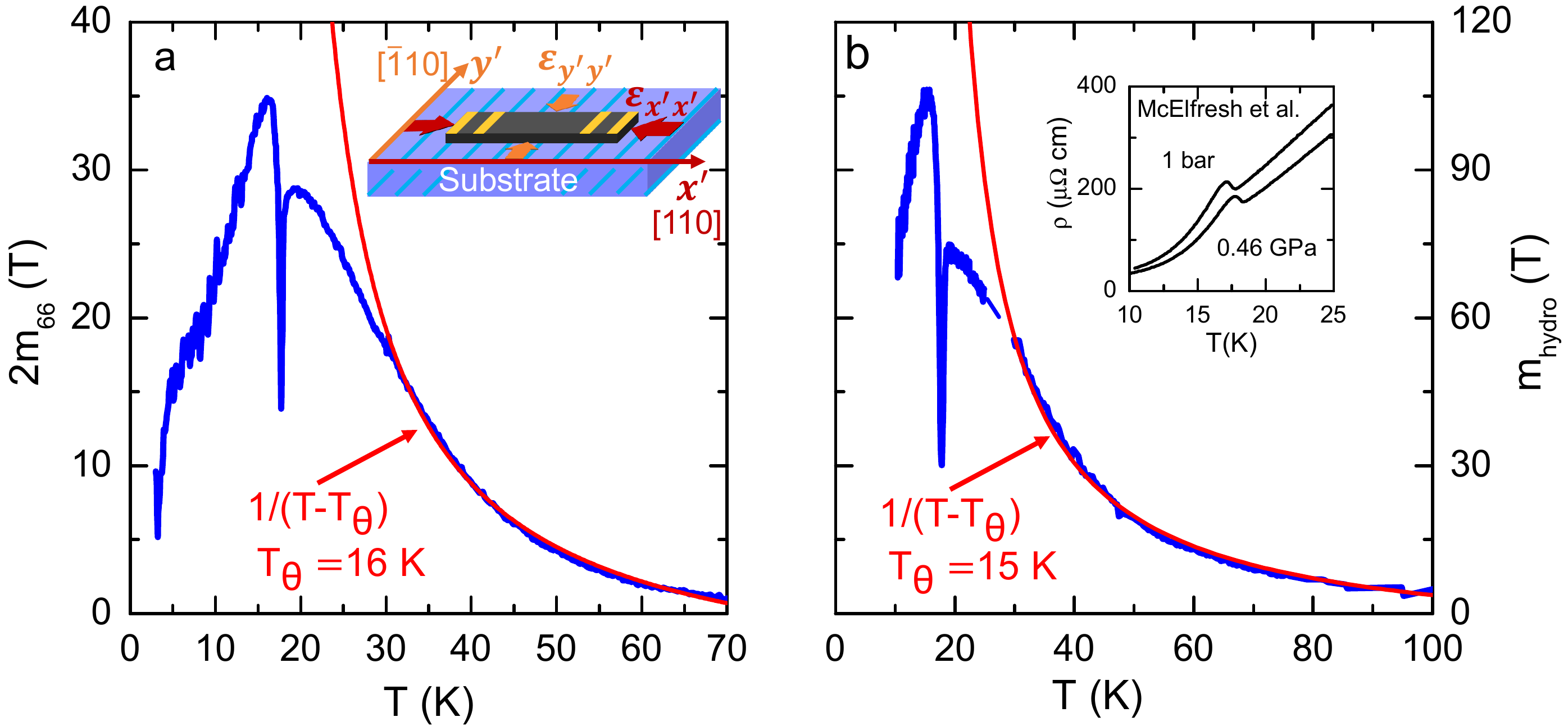}

\caption{(a) Temperature dependence of the (a) uniaxial $2m_{66}$ and (b) hydrostatic $m_{hydro}$ elastoresistivity coefficients.  The data in (a) were obtained using a uniaxial strain introduced by a glass-fiber substrate (see inset in (a) and Ref. \cite{He17_GFK_Ba122}, and  $m_{hydro}$ was calculated using the difference in resistivity between pressures of 1 bar and 0.46 GPa from McElfresh
\textit{et al.} \cite{McElfresh87_Pressure_URu2Si2} shown in the
inset of (b).\label{Fig4}}
\end{figure*}

Finally, we examine the elastoresitivity 
$2m_{66}=\frac{\rho_{x^{'}x^{'}}-\rho_{y^{'}y^{'}}}{\frac{1}{2}(\rho_{x^{'}x^{'}}+\rho_{y^{'}y^{'}})(\epsilon_{x^{'}x^{'}}-\epsilon_{y^{'}y^{'}})}$(\textcolor{black}{see Fig. \ref{Fig4}a and \cite{NoteX} for the definition of the coordinates}),
which has been argued to be a measure of the nematic susceptibility
in Fe-based superconductors \cite{ChuScience2012}. In Figure \ref{Fig4} we present our
result of $2m_{66}$ determined using a differential thermal-expansion
method \cite{He17_GFK_Ba122}. Here, the crystal was glued to a fiber-glass
substrate, which exhibits significant anisotropic thermal expansion,
so that the crystal experiences an anisotropic strain upon cooling.
The strain transmitted to the crystal is estimated by measuring
the thermal expansion of the crystal and substrate separately \cite{He17_GFK_Ba122,NoteX}.
Using the same four electrical contacts, the resistivity difference
between the crystal in free standing and strained configurations
provides the anisotropic strain response.

A sizeable strain dependence of the resistivity
is observed, with very similar $2m_{66}$ values and 
temperature dependence, as determined previously using a piezo stack
by Riggs \textit{et al.} \cite{Riggs15_Nematicity_URu2Si2}. In particular,
our data above about 30 K can also be fit with a Curie-Weiss temperature
dependence, $2m_{66}=\frac{C}{T-T_{\theta}}+2m_{66}^{0}$, with $T_{\theta}$
= 16 K which is slightly lower than $T_{HO}$ = 17.5 K, suggestive
of a sizeable nematic response. However as argued below, this suggested nematic response in elastoresistivity, which is at odds with our thermal-expansion
and Young\textquoteright s-modulus data, is most likely not due to a nematic response.

Previous work on URu$_{2}$Si$_{2}$ by McElfresh
\textit{et al.} \cite{McElfresh87_Pressure_URu2Si2} clearly demonstrate
that the resistivity is also highly sensitive to hydrostatic pressure.
One can also define a dimensionless
hydrostatic elastoresistivity coefficient as $m_{hydro}=(\frac{\Delta\rho}{\rho})/(\frac{\Delta V}{V})$
, where $\frac{\Delta\rho}{\rho}$ is the relative change of resistivity
induced by a relative change of volume $\frac{\Delta V}{V}$. To calculate
$m_{hydro}$, we took the difference in resistivity between 1 bar
and 0.46 GPa from Ref.~\cite{McElfresh87_Pressure_URu2Si2} (see inset
of Fig. \ref{Fig4}(b)), and the volume change was computed using the
bulk-modulus value of 190 GPa from Ref. \cite{Jeffries2010arXiv1002.2245J}. The inferred temperature
dependence of $m_{hydro}$ (T), plotted in Fig. \ref{Fig4}(b), remarkably
resembles that of $2m_{66}(T)$. It can also be fit by a Curie-Weiss divergence over
a similar temperature interval, and it has a similar dip at $T_{HO}$
. We find that the maximum value of $m_{hydro}(T)$ is roughly three
times larger than that of $2m_{66}$. Since there is no symmetry-breaking
strain in a hydrostatic-pressure experiment, these data clearly demonstrate
that a Curie-Weiss-like temperature dependence of an elastoresistivity
component can be obtained in the absence of nematicity, and that a
Curie-Weiss like response under a symmetry-breaking strain (as in
Fig. \ref{Fig4}(a)) does not necessarily imply that the system is
nematic. The obvious question now is what is the physics behind the
Curie-Weiss-like temperature dependence? As clearly shown in Ref. \cite{McElfresh87_Pressure_URu2Si2}, the primary effect of pressure
is to move the maximum in the resistivity, $T_{max}$, to higher temperatures.
As a consequence, the shift of $T_{max}$, together with the normalization
by $\rho(T)$ accidentally leads to a Curie-Weiss-like behavior of
$m_{hydro}(T)$. The fact that $m_{hydro}(T)\approx3\times2m_{66}(T)$
strongly suggests that $2m_{66}$ is probing a fraction of the hydrostatic part. A detailed analysis actually has shown that $2m_{66}$ and the isotropic in-plane strain effects are  nearly equal \cite{ShapiroPhD2016,NoteX}.

In summary, we find no evidence for an electronic nematic transition
associated with the HO transition in URu$_{2}$Si$_{2}$
using three sensitive experimental techniques. We thus conclude
that URu$_{2}$Si$_{2}$ does not undergo a crystallographic symmetry
reduction at $T_{HO}$, and that the HO must be restricted to tetragonal symmetry, as proposed by the recent NMR results \cite{KambePRB2015}. The HO-state is thus most likely of non-nematic rank-5 multipolar order (dotriacontapolar) order \cite{KambePRB2015,Ikeda2012}. Our results, which are consistent with several decades of intense investigations of this material \cite{Mydosh14_Review_URu2Si2}, are thus in contradiction to the recent studies reporting a symmetry reduction at the HO transition. The proponents of the nematic transition may argue
that our crystals are not clean enough, and that only the cleanest
crystals undergo this transition. A recent paper \cite{Choi_PRB2018_pressure}
found evidence for a pressure induced symmetry breaking transition near 100 K at roughly the pressure where magnetic order sets in. The purported symmetry
breaking at $T_{HO}$ found in Ref. \cite{Tonegawa14_Nematicity_URu2Si2} was included in their phase diagram and was argued to result from a reduced a-axis lattice
parameter in these crystals \cite{Choi_PRB2018_pressure}. The a-axis lattice parameters of our typical crystals (a = 4.1327(3)  Angstroms)\cite{MatsudaJPSJ2011}, on the
other hand, is larger and is thus in the stable HO region of this
phase diagram. This is advantageous, since we are most interested
in investigating the stable HO phase, and our results clearly demonstrate
that the HO phase transition in this stable region does not induce a lattice
symmetry reduction. Finally, the quantitative difference between {[}100{]} and {[}110{]} directions in elastoresistivity is intriguing, but may just
result from the natural in-equivalence of strain in {[}100{]} and {[}110{]}  directions,
which should be observable in any tetragonal system. Recent
elastoresistivity measurements on tetragonal RbFe$_{2}$As$_{2}$ and CsFe$_{2}$As$_{2}$ have
found huge elastoresistive responses with a Curie-Weiss-like temperature
dependence indicative of nematicity \cite{Ishida2018}, although previous high-resolution
thermal-expansion \cite{Hardy16_K-doped-Pnictide,AnneThesis2014} and shear-modulus measurements
 \cite{AnneThesis2014} find little evidence
of a nematic transition or nematic fluctuations in these materials.
Both RbFe$_{2}$As$_{2}$ and CsFe$_{2}$As$_{2}$, however, have large electronic Gr$\ddot{\rm u}$neiesen parameters
 at low-temperature \cite{Hardy16_K-doped-Pnictide,AnneThesis2014, Eilers16_QO_AFe2As2}, and thus 
a highly strain sensitive electronic
structure.
More
work using both thermodynamic, as well as elastoresistive
measurements, on well-characterized crystals are needed to study this
apparent discrepancy, in particular in cuprates, where
the elastoresistive response is
quite weak~\cite{Ishida2019arXiv190807167I}.

We are very thankful to I.R. Fisher for enlightening discussion about elastoresistivity measurements and interpretations. We would like to acknowledge interesting discussions with A.E. B$\ddot{\rm o}$hmer, K. Ishida, G. Knebel, T. Shibauchi, Y. Matsuda and R. Willa. L. Wang acknowledges support through  DFG  Grant No.WA4313/1-1, and K. Willa acknowledges the Humboldt Foundation for support. D. Aoki acknowledges support through KAKENHI (JP15H05882, JP15H05884, JP15K21732, JP15H05734, JP16H04006, JP19H00646)

\bibliographystyle{apsrev4-1}
\bibliography{Biblio}

\begin{thebibliography}{37}%
\makeatletter
\providecommand \@ifxundefined [1]{%
 \@ifx{#1\undefined}
}%
\providecommand \@ifnum [1]{%
 \ifnum #1\expandafter \@firstoftwo
 \else \expandafter \@secondoftwo
 \fi
}%
\providecommand \@ifx [1]{%
 \ifx #1\expandafter \@firstoftwo
 \else \expandafter \@secondoftwo
 \fi
}%
\providecommand \natexlab [1]{#1}%
\providecommand \enquote  [1]{``#1''}%
\providecommand \bibnamefont  [1]{#1}%
\providecommand \bibfnamefont [1]{#1}%
\providecommand \citenamefont [1]{#1}%
\providecommand \href@noop [0]{\@secondoftwo}%
\providecommand \href [0]{\begingroup \@sanitize@url \@href}%
\providecommand \@href[1]{\@@startlink{#1}\@@href}%
\providecommand \@@href[1]{\endgroup#1\@@endlink}%
\providecommand \@sanitize@url [0]{\catcode `\\12\catcode `\$12\catcode
  `\&12\catcode `\#12\catcode `\^12\catcode `\_12\catcode `\%12\relax}%
\providecommand \@@startlink[1]{}%
\providecommand \@@endlink[0]{}%
\providecommand \url  [0]{\begingroup\@sanitize@url \@url }%
\providecommand \@url [1]{\endgroup\@href {#1}{\urlprefix }}%
\providecommand \urlprefix  [0]{URL }%
\providecommand \Eprint [0]{\href }%
\providecommand \doibase [0]{http://dx.doi.org/}%
\providecommand \selectlanguage [0]{\@gobble}%
\providecommand \bibinfo  [0]{\@secondoftwo}%
\providecommand \bibfield  [0]{\@secondoftwo}%
\providecommand \translation [1]{[#1]}%
\providecommand \BibitemOpen [0]{}%
\providecommand \bibitemStop [0]{}%
\providecommand \bibitemNoStop [0]{.\EOS\space}%
\providecommand \EOS [0]{\spacefactor3000\relax}%
\providecommand \BibitemShut  [1]{\csname bibitem#1\endcsname}%
\let\auto@bib@innerbib\@empty
\bibitem [{\citenamefont {Mydosh}\ and\ \citenamefont
  {Oppeneer}(2011)}]{Mydosh11_review_URu2Si2}%
  \BibitemOpen
  \bibfield  {author} {\bibinfo {author} {\bibfnamefont {J.~A.}\ \bibnamefont
  {Mydosh}}\ and\ \bibinfo {author} {\bibfnamefont {P.~M.}\ \bibnamefont
  {Oppeneer}},\ }\href {\doibase 10.1103/RevModPhys.83.1301} {\bibfield
  {journal} {\bibinfo  {journal} {Rev. Mod. Phys.}\ }\textbf {\bibinfo {volume}
  {83}},\ \bibinfo {pages} {1301} (\bibinfo {year} {2011})}\BibitemShut
  {NoStop}%
\bibitem [{\citenamefont {Mydosh}\ and\ \citenamefont
  {Oppeneer}(2014)}]{Mydosh14_Review_URu2Si2}%
  \BibitemOpen
  \bibfield  {author} {\bibinfo {author} {\bibfnamefont {J.}~\bibnamefont
  {Mydosh}}\ and\ \bibinfo {author} {\bibfnamefont {P.}~\bibnamefont
  {Oppeneer}},\ }\href {\doibase 10.1080/14786435.2014.916428} {\bibfield
  {journal} {\bibinfo  {journal} {Philosophical Magazine}\ }\textbf {\bibinfo
  {volume} {94}},\ \bibinfo {pages} {3642} (\bibinfo {year}
  {2014})}\BibitemShut {NoStop}%
\bibitem [{\citenamefont {Okazaki}\ \emph {et~al.}(2011)\citenamefont
  {Okazaki}, \citenamefont {Shibauchi}, \citenamefont {Shi}, \citenamefont
  {Haga}, \citenamefont {Matsuda}, \citenamefont {Yamamoto}, \citenamefont
  {Onuki}, \citenamefont {Ikeda},\ and\ \citenamefont
  {Matsuda}}]{Okazaki11_Torque_URu2Si2}%
  \BibitemOpen
  \bibfield  {author} {\bibinfo {author} {\bibfnamefont {R.}~\bibnamefont
  {Okazaki}}, \bibinfo {author} {\bibfnamefont {T.}~\bibnamefont {Shibauchi}},
  \bibinfo {author} {\bibfnamefont {H.~J.}\ \bibnamefont {Shi}}, \bibinfo
  {author} {\bibfnamefont {Y.}~\bibnamefont {Haga}}, \bibinfo {author}
  {\bibfnamefont {T.~D.}\ \bibnamefont {Matsuda}}, \bibinfo {author}
  {\bibfnamefont {E.}~\bibnamefont {Yamamoto}}, \bibinfo {author}
  {\bibfnamefont {Y.}~\bibnamefont {Onuki}}, \bibinfo {author} {\bibfnamefont
  {H.}~\bibnamefont {Ikeda}}, \ and\ \bibinfo {author} {\bibfnamefont
  {Y.}~\bibnamefont {Matsuda}},\ }\href {\doibase 10.1126/science.1197358}
  {\bibfield  {journal} {\bibinfo  {journal} {Science}\ }\textbf {\bibinfo
  {volume} {331}},\ \bibinfo {pages} {439} (\bibinfo {year}
  {2011})}\BibitemShut {NoStop}%
\bibitem [{\citenamefont {Tonegawa}\ \emph {et~al.}(2014)\citenamefont
  {Tonegawa}, \citenamefont {Kasahara}, \citenamefont {Fukuda}, \citenamefont
  {Sugimoto}, \citenamefont {Yasuda}, \citenamefont {Tsuruhara}, \citenamefont
  {Watanabe}, \citenamefont {Mizukami}, \citenamefont {Haga}, \citenamefont
  {Matsuda}, \citenamefont {Yamamoto}, \citenamefont {Onuki}, \citenamefont
  {Ikeda}, \citenamefont {Matsuda},\ and\ \citenamefont
  {Shibauchi}}]{Tonegawa14_Nematicity_URu2Si2}%
  \BibitemOpen
  \bibfield  {author} {\bibinfo {author} {\bibfnamefont {S.}~\bibnamefont
  {Tonegawa}}, \bibinfo {author} {\bibfnamefont {S.}~\bibnamefont {Kasahara}},
  \bibinfo {author} {\bibfnamefont {T.}~\bibnamefont {Fukuda}}, \bibinfo
  {author} {\bibfnamefont {K.}~\bibnamefont {Sugimoto}}, \bibinfo {author}
  {\bibfnamefont {N.}~\bibnamefont {Yasuda}}, \bibinfo {author} {\bibfnamefont
  {Y.}~\bibnamefont {Tsuruhara}}, \bibinfo {author} {\bibfnamefont
  {D.}~\bibnamefont {Watanabe}}, \bibinfo {author} {\bibfnamefont
  {Y.}~\bibnamefont {Mizukami}}, \bibinfo {author} {\bibfnamefont
  {Y.}~\bibnamefont {Haga}}, \bibinfo {author} {\bibfnamefont {T.~D.}\
  \bibnamefont {Matsuda}}, \bibinfo {author} {\bibfnamefont {E.}~\bibnamefont
  {Yamamoto}}, \bibinfo {author} {\bibfnamefont {Y.}~\bibnamefont {Onuki}},
  \bibinfo {author} {\bibfnamefont {H.}~\bibnamefont {Ikeda}}, \bibinfo
  {author} {\bibfnamefont {Y.}~\bibnamefont {Matsuda}}, \ and\ \bibinfo
  {author} {\bibfnamefont {T.}~\bibnamefont {Shibauchi}},\ }\href
  {https://doi.org/10.1038/ncomms5188} {\bibfield  {journal} {\bibinfo
  {journal} {Nature Communications}\ }\textbf {\bibinfo {volume} {5}},\
  \bibinfo {pages} {4188 EP } (\bibinfo {year} {2014})}\BibitemShut {NoStop}%
\bibitem [{\citenamefont {Riggs}\ \emph {et~al.}(2015)\citenamefont {Riggs},
  \citenamefont {Shapiro}, \citenamefont {Maharaj}, \citenamefont {Raghu},
  \citenamefont {Bauer}, \citenamefont {Baumbach}, \citenamefont
  {Giraldo-Gallo}, \citenamefont {Wartenbe},\ and\ \citenamefont
  {Fisher}}]{Riggs15_Nematicity_URu2Si2}%
  \BibitemOpen
  \bibfield  {author} {\bibinfo {author} {\bibfnamefont {S.~C.}\ \bibnamefont
  {Riggs}}, \bibinfo {author} {\bibfnamefont {M.~C.}\ \bibnamefont {Shapiro}},
  \bibinfo {author} {\bibfnamefont {A.~V.}\ \bibnamefont {Maharaj}}, \bibinfo
  {author} {\bibfnamefont {S.}~\bibnamefont {Raghu}}, \bibinfo {author}
  {\bibfnamefont {E.~D.}\ \bibnamefont {Bauer}}, \bibinfo {author}
  {\bibfnamefont {R.~E.}\ \bibnamefont {Baumbach}}, \bibinfo {author}
  {\bibfnamefont {P.}~\bibnamefont {Giraldo-Gallo}}, \bibinfo {author}
  {\bibfnamefont {M.}~\bibnamefont {Wartenbe}}, \ and\ \bibinfo {author}
  {\bibfnamefont {I.~R.}\ \bibnamefont {Fisher}},\ }\href
  {https://doi.org/10.1038/ncomms7425} {\bibfield  {journal} {\bibinfo
  {journal} {Nature Communications}\ }\textbf {\bibinfo {volume} {6}},\
  \bibinfo {pages} {6425 EP } (\bibinfo {year} {2015})}\BibitemShut {NoStop}%
\bibitem [{\citenamefont {Ikeda}\ \emph {et~al.}(2012)\citenamefont {Ikeda},
  \citenamefont {Suzuki}, \citenamefont {Arita}, \citenamefont {Takimoto},
  \citenamefont {Shibauchi},\ and\ \citenamefont {Matsuda}}]{Ikeda2012}%
  \BibitemOpen
  \bibfield  {author} {\bibinfo {author} {\bibfnamefont {H.}~\bibnamefont
  {Ikeda}}, \bibinfo {author} {\bibfnamefont {M.-T.}\ \bibnamefont {Suzuki}},
  \bibinfo {author} {\bibfnamefont {R.}~\bibnamefont {Arita}}, \bibinfo
  {author} {\bibfnamefont {T.}~\bibnamefont {Takimoto}}, \bibinfo {author}
  {\bibfnamefont {T.}~\bibnamefont {Shibauchi}}, \ and\ \bibinfo {author}
  {\bibfnamefont {Y.}~\bibnamefont {Matsuda}},\ }\href {\doibase
  10.1038/nphys2330} {\bibfield  {journal} {\bibinfo  {journal} {Nature
  Physics}\ }\textbf {\bibinfo {volume} {8}},\ \bibinfo {pages} {528} (\bibinfo
  {year} {2012})}\BibitemShut {NoStop}%
\bibitem [{\citenamefont {{Tabata}}\ \emph {et~al.}(2014)\citenamefont
  {{Tabata}}, \citenamefont {{Inami}}, \citenamefont {{Michimura}},
  \citenamefont {{Yokoyama}}, \citenamefont {{Hidaka}}, \citenamefont
  {{Yanagisawa}},\ and\ \citenamefont {{Amitsuka}}}]{2014tabata}%
  \BibitemOpen
  \bibfield  {author} {\bibinfo {author} {\bibfnamefont {C.}~\bibnamefont
  {{Tabata}}}, \bibinfo {author} {\bibfnamefont {T.}~\bibnamefont {{Inami}}},
  \bibinfo {author} {\bibfnamefont {S.}~\bibnamefont {{Michimura}}}, \bibinfo
  {author} {\bibfnamefont {M.}~\bibnamefont {{Yokoyama}}}, \bibinfo {author}
  {\bibfnamefont {H.}~\bibnamefont {{Hidaka}}}, \bibinfo {author}
  {\bibfnamefont {T.}~\bibnamefont {{Yanagisawa}}}, \ and\ \bibinfo {author}
  {\bibfnamefont {H.}~\bibnamefont {{Amitsuka}}},\ }\href {\doibase
  10.1080/14786435.2014.952701} {\bibfield  {journal} {\bibinfo  {journal}
  {Philosophical Magazine}\ }\textbf {\bibinfo {volume} {94}},\ \bibinfo
  {pages} {3691} (\bibinfo {year} {2014})}\BibitemShut {NoStop}%
\bibitem [{\citenamefont {Kambe}\ \emph {et~al.}(2018)\citenamefont {Kambe},
  \citenamefont {Tokunaga}, \citenamefont {Sakai}, \citenamefont {Hattori},
  \citenamefont {Higa}, \citenamefont {Matsuda}, \citenamefont {Haga},
  \citenamefont {Walstedt},\ and\ \citenamefont {Harima}}]{KambePRB97NMR}%
  \BibitemOpen
  \bibfield  {author} {\bibinfo {author} {\bibfnamefont {S.}~\bibnamefont
  {Kambe}}, \bibinfo {author} {\bibfnamefont {Y.}~\bibnamefont {Tokunaga}},
  \bibinfo {author} {\bibfnamefont {H.}~\bibnamefont {Sakai}}, \bibinfo
  {author} {\bibfnamefont {T.}~\bibnamefont {Hattori}}, \bibinfo {author}
  {\bibfnamefont {N.}~\bibnamefont {Higa}}, \bibinfo {author} {\bibfnamefont
  {T.~D.}\ \bibnamefont {Matsuda}}, \bibinfo {author} {\bibfnamefont
  {Y.}~\bibnamefont {Haga}}, \bibinfo {author} {\bibfnamefont {R.~E.}\
  \bibnamefont {Walstedt}}, \ and\ \bibinfo {author} {\bibfnamefont
  {H.}~\bibnamefont {Harima}},\ }\href {\doibase 10.1103/PhysRevB.97.235142}
  {\bibfield  {journal} {\bibinfo  {journal} {Phys. Rev. B}\ }\textbf {\bibinfo
  {volume} {97}},\ \bibinfo {pages} {235142} (\bibinfo {year}
  {2018})}\BibitemShut {NoStop}%
\bibitem [{\citenamefont {Kambe}\ \emph {et~al.}(2015)\citenamefont {Kambe},
  \citenamefont {Tokunaga}, \citenamefont {Sakai},\ and\ \citenamefont
  {Walstedt}}]{KambePRB2015}%
  \BibitemOpen
  \bibfield  {author} {\bibinfo {author} {\bibfnamefont {S.}~\bibnamefont
  {Kambe}}, \bibinfo {author} {\bibfnamefont {Y.}~\bibnamefont {Tokunaga}},
  \bibinfo {author} {\bibfnamefont {H.}~\bibnamefont {Sakai}}, \ and\ \bibinfo
  {author} {\bibfnamefont {R.~E.}\ \bibnamefont {Walstedt}},\ }\href {\doibase
  10.1103/PhysRevB.91.035111} {\bibfield  {journal} {\bibinfo  {journal} {Phys.
  Rev. B}\ }\textbf {\bibinfo {volume} {91}},\ \bibinfo {pages} {035111}
  (\bibinfo {year} {2015})}\BibitemShut {NoStop}%
\bibitem [{\citenamefont {Bourdarot}\ \emph {et~al.}(2010)\citenamefont
  {Bourdarot}, \citenamefont {Hassinger}, \citenamefont {Raymond},
  \citenamefont {Aoki}, \citenamefont {Taufour}, \citenamefont {Regnault},\
  and\ \citenamefont {Flouquet}}]{BourdarotJPSJ2010}%
  \BibitemOpen
  \bibfield  {author} {\bibinfo {author} {\bibfnamefont {F.}~\bibnamefont
  {Bourdarot}}, \bibinfo {author} {\bibfnamefont {E.}~\bibnamefont
  {Hassinger}}, \bibinfo {author} {\bibfnamefont {S.}~\bibnamefont {Raymond}},
  \bibinfo {author} {\bibfnamefont {D.}~\bibnamefont {Aoki}}, \bibinfo {author}
  {\bibfnamefont {V.}~\bibnamefont {Taufour}}, \bibinfo {author} {\bibfnamefont
  {L.-P.}\ \bibnamefont {Regnault}}, \ and\ \bibinfo {author} {\bibfnamefont
  {J.}~\bibnamefont {Flouquet}},\ }\href {\doibase 10.1143/JPSJ.79.064719}
  {\bibfield  {journal} {\bibinfo  {journal} {Journal of the Physical Society
  of Japan}\ }\textbf {\bibinfo {volume} {79}},\ \bibinfo {pages} {064719}
  (\bibinfo {year} {2010})}\BibitemShut {NoStop}%
\bibitem [{\citenamefont {B\"ohmer}\ \emph {et~al.}(2015)\citenamefont
  {B\"ohmer}, \citenamefont {Arai}, \citenamefont {Hardy}, \citenamefont
  {Hattori}, \citenamefont {Iye}, \citenamefont {Wolf}, \citenamefont
  {L\"ohneysen}, \citenamefont {Ishida},\ and\ \citenamefont
  {Meingast}}]{Anna_FeSe_PRL2015}%
  \BibitemOpen
  \bibfield  {author} {\bibinfo {author} {\bibfnamefont {A.~E.}\ \bibnamefont
  {B\"ohmer}}, \bibinfo {author} {\bibfnamefont {T.}~\bibnamefont {Arai}},
  \bibinfo {author} {\bibfnamefont {F.}~\bibnamefont {Hardy}}, \bibinfo
  {author} {\bibfnamefont {T.}~\bibnamefont {Hattori}}, \bibinfo {author}
  {\bibfnamefont {T.}~\bibnamefont {Iye}}, \bibinfo {author} {\bibfnamefont
  {T.}~\bibnamefont {Wolf}}, \bibinfo {author} {\bibfnamefont {H.~v.}\
  \bibnamefont {L\"ohneysen}}, \bibinfo {author} {\bibfnamefont
  {K.}~\bibnamefont {Ishida}}, \ and\ \bibinfo {author} {\bibfnamefont
  {C.}~\bibnamefont {Meingast}},\ }\href {\doibase
  10.1103/PhysRevLett.114.027001} {\bibfield  {journal} {\bibinfo  {journal}
  {Phys. Rev. Lett.}\ }\textbf {\bibinfo {volume} {114}},\ \bibinfo {pages}
  {027001} (\bibinfo {year} {2015})}\BibitemShut {NoStop}%
\bibitem [{\citenamefont {Wang}\ \emph {et~al.}(2018)\citenamefont {Wang},
  \citenamefont {He}, \citenamefont {Hardy}, \citenamefont {Adelmann},
  \citenamefont {Wolf}, \citenamefont {Merz}, \citenamefont {Schweiss},\ and\
  \citenamefont {Meingast}}]{Wang18_Nematicity_C4_Phase}%
  \BibitemOpen
  \bibfield  {author} {\bibinfo {author} {\bibfnamefont {L.}~\bibnamefont
  {Wang}}, \bibinfo {author} {\bibfnamefont {M.}~\bibnamefont {He}}, \bibinfo
  {author} {\bibfnamefont {F.}~\bibnamefont {Hardy}}, \bibinfo {author}
  {\bibfnamefont {P.}~\bibnamefont {Adelmann}}, \bibinfo {author}
  {\bibfnamefont {T.}~\bibnamefont {Wolf}}, \bibinfo {author} {\bibfnamefont
  {M.}~\bibnamefont {Merz}}, \bibinfo {author} {\bibfnamefont {P.}~\bibnamefont
  {Schweiss}}, \ and\ \bibinfo {author} {\bibfnamefont {C.}~\bibnamefont
  {Meingast}},\ }\href {\doibase 10.1103/PhysRevB.97.224518} {\bibfield
  {journal} {\bibinfo  {journal} {Phys. Rev. B}\ }\textbf {\bibinfo {volume}
  {97}},\ \bibinfo {pages} {224518} (\bibinfo {year} {2018})}\BibitemShut
  {NoStop}%
\bibitem [{\citenamefont {Wang}\ \emph {et~al.}(2019)\citenamefont {Wang},
  \citenamefont {He}, \citenamefont {Scherer}, \citenamefont {Hardy},
  \citenamefont {Schweiss}, \citenamefont {Wolf}, \citenamefont {Merz},
  \citenamefont {Andersen},\ and\ \citenamefont {Meingast}}]{Liran_JPSJ2019}%
  \BibitemOpen
  \bibfield  {author} {\bibinfo {author} {\bibfnamefont {L.}~\bibnamefont
  {Wang}}, \bibinfo {author} {\bibfnamefont {M.}~\bibnamefont {He}}, \bibinfo
  {author} {\bibfnamefont {D.~D.}\ \bibnamefont {Scherer}}, \bibinfo {author}
  {\bibfnamefont {F.}~\bibnamefont {Hardy}}, \bibinfo {author} {\bibfnamefont
  {P.}~\bibnamefont {Schweiss}}, \bibinfo {author} {\bibfnamefont
  {T.}~\bibnamefont {Wolf}}, \bibinfo {author} {\bibfnamefont {M.}~\bibnamefont
  {Merz}}, \bibinfo {author} {\bibfnamefont {B.~M.}\ \bibnamefont {Andersen}},
  \ and\ \bibinfo {author} {\bibfnamefont {C.}~\bibnamefont {Meingast}},\
  }\href {\doibase 10.7566/JPSJ.88.104710} {\bibfield  {journal} {\bibinfo
  {journal} {Journal of the Physical Society of Japan}\ }\textbf {\bibinfo
  {volume} {88}},\ \bibinfo {pages} {104710} (\bibinfo {year}
  {2019})}\BibitemShut {NoStop}%
\bibitem [{\citenamefont {B\"ohmer}\ \emph {et~al.}(2014)\citenamefont
  {B\"ohmer}, \citenamefont {Burger}, \citenamefont {Hardy}, \citenamefont
  {Wolf}, \citenamefont {Schweiss}, \citenamefont {Fromknecht}, \citenamefont
  {Reinecker}, \citenamefont {Schranz},\ and\ \citenamefont
  {Meingast}}]{Boehmer14_Nematic_Susceptibility}%
  \BibitemOpen
  \bibfield  {author} {\bibinfo {author} {\bibfnamefont {A.~E.}\ \bibnamefont
  {B\"ohmer}}, \bibinfo {author} {\bibfnamefont {P.}~\bibnamefont {Burger}},
  \bibinfo {author} {\bibfnamefont {F.}~\bibnamefont {Hardy}}, \bibinfo
  {author} {\bibfnamefont {T.}~\bibnamefont {Wolf}}, \bibinfo {author}
  {\bibfnamefont {P.}~\bibnamefont {Schweiss}}, \bibinfo {author}
  {\bibfnamefont {R.}~\bibnamefont {Fromknecht}}, \bibinfo {author}
  {\bibfnamefont {M.}~\bibnamefont {Reinecker}}, \bibinfo {author}
  {\bibfnamefont {W.}~\bibnamefont {Schranz}}, \ and\ \bibinfo {author}
  {\bibfnamefont {C.}~\bibnamefont {Meingast}},\ }\href {\doibase
  10.1103/PhysRevLett.112.047001} {\bibfield  {journal} {\bibinfo  {journal}
  {Phys. Rev. Lett.}\ }\textbf {\bibinfo {volume} {112}},\ \bibinfo {pages}
  {047001} (\bibinfo {year} {2014})}\BibitemShut {NoStop}%
\bibitem [{\citenamefont {B{\"o}hmer}\ and\ \citenamefont
  {Meingast}(2016)}]{Boehmer16_Nematicity_CR}%
  \BibitemOpen
  \bibfield  {author} {\bibinfo {author} {\bibfnamefont {A.~E.}\ \bibnamefont
  {B{\"o}hmer}}\ and\ \bibinfo {author} {\bibfnamefont {C.}~\bibnamefont
  {Meingast}},\ }\href {\doibase https://doi.org/10.1016/j.crhy.2015.07.001}
  {\bibfield  {journal} {\bibinfo  {journal} {Comptes Rendus Physique}\
  }\textbf {\bibinfo {volume} {17}},\ \bibinfo {pages} {90 } (\bibinfo {year}
  {2016})}\BibitemShut {NoStop}%
\bibitem [{\citenamefont {Chu}\ \emph {et~al.}(2012)\citenamefont {Chu},
  \citenamefont {Kuo}, \citenamefont {Analytis},\ and\ \citenamefont
  {Fisher}}]{ChuScience2012}%
  \BibitemOpen
  \bibfield  {author} {\bibinfo {author} {\bibfnamefont {J.-H.}\ \bibnamefont
  {Chu}}, \bibinfo {author} {\bibfnamefont {H.-H.}\ \bibnamefont {Kuo}},
  \bibinfo {author} {\bibfnamefont {J.~G.}\ \bibnamefont {Analytis}}, \ and\
  \bibinfo {author} {\bibfnamefont {I.~R.}\ \bibnamefont {Fisher}},\ }\href
  {\doibase 10.1126/science.1221713} {\bibfield  {journal} {\bibinfo  {journal}
  {Science}\ }\textbf {\bibinfo {volume} {337}},\ \bibinfo {pages} {710}
  (\bibinfo {year} {2012})}\BibitemShut {NoStop}%
\bibitem [{\citenamefont {Aoki}\ \emph {et~al.}(2010)\citenamefont {Aoki},
  \citenamefont {Bourdarot}, \citenamefont {Hassinger}, \citenamefont {Knebel},
  \citenamefont {Miyake}, \citenamefont {Raymond}, \citenamefont {Taufour},\
  and\ \citenamefont {Flouquet}}]{Aoki10_Reentrance_U122}%
  \BibitemOpen
  \bibfield  {author} {\bibinfo {author} {\bibfnamefont {D.}~\bibnamefont
  {Aoki}}, \bibinfo {author} {\bibfnamefont {F.}~\bibnamefont {Bourdarot}},
  \bibinfo {author} {\bibfnamefont {E.}~\bibnamefont {Hassinger}}, \bibinfo
  {author} {\bibfnamefont {G.}~\bibnamefont {Knebel}}, \bibinfo {author}
  {\bibfnamefont {A.}~\bibnamefont {Miyake}}, \bibinfo {author} {\bibfnamefont
  {S.}~\bibnamefont {Raymond}}, \bibinfo {author} {\bibfnamefont
  {V.}~\bibnamefont {Taufour}}, \ and\ \bibinfo {author} {\bibfnamefont
  {J.}~\bibnamefont {Flouquet}},\ }\href
  {http://stacks.iop.org/0953-8984/22/i=16/a=164205} {\bibfield  {journal}
  {\bibinfo  {journal} {Journal of Physics: Condensed Matter}\ }\textbf
  {\bibinfo {volume} {22}},\ \bibinfo {pages} {164205} (\bibinfo {year}
  {2010})}\BibitemShut {NoStop}%
\bibitem [{\citenamefont {D.~Matsuda}\ \emph {et~al.}(2011)\citenamefont
  {D.~Matsuda}, \citenamefont {Hassinger}, \citenamefont {Aoki}, \citenamefont
  {Taufour}, \citenamefont {Knebel}, \citenamefont {Tateiwa}, \citenamefont
  {Yamamoto}, \citenamefont {Haga}, \citenamefont {Onuki}, \citenamefont
  {Fisk},\ and\ \citenamefont {Flouquet}}]{MatsudaJPSJ2011}%
  \BibitemOpen
  \bibfield  {author} {\bibinfo {author} {\bibfnamefont {T.}~\bibnamefont
  {D.~Matsuda}}, \bibinfo {author} {\bibfnamefont {E.}~\bibnamefont
  {Hassinger}}, \bibinfo {author} {\bibfnamefont {D.}~\bibnamefont {Aoki}},
  \bibinfo {author} {\bibfnamefont {V.}~\bibnamefont {Taufour}}, \bibinfo
  {author} {\bibfnamefont {G.}~\bibnamefont {Knebel}}, \bibinfo {author}
  {\bibfnamefont {N.}~\bibnamefont {Tateiwa}}, \bibinfo {author} {\bibfnamefont
  {E.}~\bibnamefont {Yamamoto}}, \bibinfo {author} {\bibfnamefont
  {Y.}~\bibnamefont {Haga}}, \bibinfo {author} {\bibfnamefont {Y.}~\bibnamefont
  {Onuki}}, \bibinfo {author} {\bibfnamefont {Z.}~\bibnamefont {Fisk}}, \ and\
  \bibinfo {author} {\bibfnamefont {J.}~\bibnamefont {Flouquet}},\ }\href
  {\doibase 10.1143/JPSJ.80.114710} {\bibfield  {journal} {\bibinfo  {journal}
  {Journal of the Physical Society of Japan}\ }\textbf {\bibinfo {volume}
  {80}},\ \bibinfo {pages} {114710} (\bibinfo {year} {2011})}\BibitemShut
  {NoStop}%
\bibitem [{\citenamefont {Meingast}\ \emph {et~al.}(1990)\citenamefont
  {Meingast}, \citenamefont {Blank}, \citenamefont {B\"urkle}, \citenamefont
  {Obst}, \citenamefont {Wolf}, \citenamefont {W\"uhl}, \citenamefont
  {Selvamanickam},\ and\ \citenamefont
  {Salama}}]{Meingast91_Thermal-Epansion_YBCO}%
  \BibitemOpen
  \bibfield  {author} {\bibinfo {author} {\bibfnamefont {C.}~\bibnamefont
  {Meingast}}, \bibinfo {author} {\bibfnamefont {B.}~\bibnamefont {Blank}},
  \bibinfo {author} {\bibfnamefont {H.}~\bibnamefont {B\"urkle}}, \bibinfo
  {author} {\bibfnamefont {B.}~\bibnamefont {Obst}}, \bibinfo {author}
  {\bibfnamefont {T.}~\bibnamefont {Wolf}}, \bibinfo {author} {\bibfnamefont
  {H.}~\bibnamefont {W\"uhl}}, \bibinfo {author} {\bibfnamefont
  {V.}~\bibnamefont {Selvamanickam}}, \ and\ \bibinfo {author} {\bibfnamefont
  {K.}~\bibnamefont {Salama}},\ }\href {\doibase 10.1103/PhysRevB.41.11299}
  {\bibfield  {journal} {\bibinfo  {journal} {Phys. Rev. B}\ }\textbf {\bibinfo
  {volume} {41}},\ \bibinfo {pages} {11299} (\bibinfo {year}
  {1990})}\BibitemShut {NoStop}%
\bibitem [{\citenamefont {Marcenat}(1986)}]{Marcenat_Thesis}%
  \BibitemOpen
  \bibfield  {author} {\bibinfo {author} {\bibfnamefont {C.}~\bibnamefont
  {Marcenat}},\ }\emph {\bibinfo {title} {\textup{Etude calorim{\'e}trique sous
  champ magn{\'e}tique des phases basses temp{\'e}ratures des compos{\'e}s
  Kondo ordonn{\'e}s: CeB$_{6}$ et TmS}}},\ \href@noop {} {Ph.D. thesis},\
  \bibinfo  {school} {Universit{\'e} scientifique et m{\'e}dicale de Grenoble}
  (\bibinfo {year} {1986})\BibitemShut {NoStop}%
\bibitem [{\citenamefont {Riegel}\ and\ \citenamefont
  {Weber}(1986)}]{Riegel86_Dual_Slope_SHC}%
  \BibitemOpen
  \bibfield  {author} {\bibinfo {author} {\bibfnamefont {S.}~\bibnamefont
  {Riegel}}\ and\ \bibinfo {author} {\bibfnamefont {G.}~\bibnamefont {Weber}},\
  }\href {http://stacks.iop.org/0022-3735/19/i=10/a=006} {\bibfield  {journal}
  {\bibinfo  {journal} {Journal of Physics E: Scientific Instruments}\ }\textbf
  {\bibinfo {volume} {19}},\ \bibinfo {pages} {790} (\bibinfo {year}
  {1986})}\BibitemShut {NoStop}%
\bibitem [{\citenamefont {He}\ \emph {et~al.}(2017)\citenamefont {He},
  \citenamefont {Wang}, \citenamefont {Ahn}, \citenamefont {Hardy},
  \citenamefont {Wolf}, \citenamefont {Adelmann}, \citenamefont {Schmalian},
  \citenamefont {Eremin},\ and\ \citenamefont {Meingast}}]{He17_GFK_Ba122}%
  \BibitemOpen
  \bibfield  {author} {\bibinfo {author} {\bibfnamefont {M.}~\bibnamefont
  {He}}, \bibinfo {author} {\bibfnamefont {L.}~\bibnamefont {Wang}}, \bibinfo
  {author} {\bibfnamefont {F.}~\bibnamefont {Ahn}}, \bibinfo {author}
  {\bibfnamefont {F.}~\bibnamefont {Hardy}}, \bibinfo {author} {\bibfnamefont
  {T.}~\bibnamefont {Wolf}}, \bibinfo {author} {\bibfnamefont {P.}~\bibnamefont
  {Adelmann}}, \bibinfo {author} {\bibfnamefont {J.}~\bibnamefont {Schmalian}},
  \bibinfo {author} {\bibfnamefont {I.}~\bibnamefont {Eremin}}, \ and\ \bibinfo
  {author} {\bibfnamefont {C.}~\bibnamefont {Meingast}},\ }\href {\doibase
  10.1038/s41467-017-00712-3} {\bibfield  {journal} {\bibinfo  {journal}
  {Nature Communications}\ }\textbf {\bibinfo {volume} {8}},\ \bibinfo {pages}
  {504} (\bibinfo {year} {2017})}\BibitemShut {NoStop}%
\bibitem [{Not()}]{NoteX}%
  \BibitemOpen
  \href@noop {} {}\bibinfo {note} {Supplemental Material}\BibitemShut {NoStop}%
\bibitem [{\citenamefont {B\"ohmer}\ \emph {et~al.}(2013)\citenamefont
  {B\"ohmer}, \citenamefont {Hardy}, \citenamefont {Eilers}, \citenamefont
  {Ernst}, \citenamefont {Adelmann}, \citenamefont {Schweiss}, \citenamefont
  {Wolf},\ and\ \citenamefont {Meingast}}]{Boehmer13_FeSe_Growth}%
  \BibitemOpen
  \bibfield  {author} {\bibinfo {author} {\bibfnamefont {A.~E.}\ \bibnamefont
  {B\"ohmer}}, \bibinfo {author} {\bibfnamefont {F.}~\bibnamefont {Hardy}},
  \bibinfo {author} {\bibfnamefont {F.}~\bibnamefont {Eilers}}, \bibinfo
  {author} {\bibfnamefont {D.}~\bibnamefont {Ernst}}, \bibinfo {author}
  {\bibfnamefont {P.}~\bibnamefont {Adelmann}}, \bibinfo {author}
  {\bibfnamefont {P.}~\bibnamefont {Schweiss}}, \bibinfo {author}
  {\bibfnamefont {T.}~\bibnamefont {Wolf}}, \ and\ \bibinfo {author}
  {\bibfnamefont {C.}~\bibnamefont {Meingast}},\ }\href {\doibase
  10.1103/PhysRevB.87.180505} {\bibfield  {journal} {\bibinfo  {journal} {Phys.
  Rev. B}\ }\textbf {\bibinfo {volume} {87}},\ \bibinfo {pages} {180505}
  (\bibinfo {year} {2013})}\BibitemShut {NoStop}%
\bibitem [{\citenamefont {B{\"o}hmer}\ \emph {et~al.}(2015)\citenamefont
  {B{\"o}hmer}, \citenamefont {Hardy}, \citenamefont {Wang}, \citenamefont
  {Wolf}, \citenamefont {Schweiss},\ and\ \citenamefont
  {Meingast}}]{Boehmer15_C4phase_BaK122}%
  \BibitemOpen
  \bibfield  {author} {\bibinfo {author} {\bibfnamefont {A.~E.}\ \bibnamefont
  {B{\"o}hmer}}, \bibinfo {author} {\bibfnamefont {F.}~\bibnamefont {Hardy}},
  \bibinfo {author} {\bibfnamefont {L.}~\bibnamefont {Wang}}, \bibinfo {author}
  {\bibfnamefont {T.}~\bibnamefont {Wolf}}, \bibinfo {author} {\bibfnamefont
  {P.}~\bibnamefont {Schweiss}}, \ and\ \bibinfo {author} {\bibfnamefont
  {C.}~\bibnamefont {Meingast}},\ }\href {https://doi.org/10.1038/ncomms8911}
  {\bibfield  {journal} {\bibinfo  {journal} {Nature Communications}\ }\textbf
  {\bibinfo {volume} {6}},\ \bibinfo {pages} {7911 EP } (\bibinfo {year}
  {2015})}\BibitemShut {NoStop}%
\bibitem [{\citenamefont {Wang}\ \emph {et~al.}(2016)\citenamefont {Wang},
  \citenamefont {Hardy}, \citenamefont {B\"ohmer}, \citenamefont {Wolf},
  \citenamefont {Schweiss},\ and\ \citenamefont
  {Meingast}}]{Wang16_Entropy_C4_phase}%
  \BibitemOpen
  \bibfield  {author} {\bibinfo {author} {\bibfnamefont {L.}~\bibnamefont
  {Wang}}, \bibinfo {author} {\bibfnamefont {F.}~\bibnamefont {Hardy}},
  \bibinfo {author} {\bibfnamefont {A.~E.}\ \bibnamefont {B\"ohmer}}, \bibinfo
  {author} {\bibfnamefont {T.}~\bibnamefont {Wolf}}, \bibinfo {author}
  {\bibfnamefont {P.}~\bibnamefont {Schweiss}}, \ and\ \bibinfo {author}
  {\bibfnamefont {C.}~\bibnamefont {Meingast}},\ }\href {\doibase
  10.1103/PhysRevB.93.014514} {\bibfield  {journal} {\bibinfo  {journal} {Phys.
  Rev. B}\ }\textbf {\bibinfo {volume} {93}},\ \bibinfo {pages} {014514}
  (\bibinfo {year} {2016})}\BibitemShut {NoStop}%
\bibitem [{\citenamefont {de~Visser}\ \emph {et~al.}(1986)\citenamefont
  {de~Visser}, \citenamefont {Kayzel}, \citenamefont {Menovsky}, \citenamefont
  {Franse}, \citenamefont {van~den Berg},\ and\ \citenamefont
  {Nieuwenhuys}}]{deVisser86_Alpha_URu2Si2}%
  \BibitemOpen
  \bibfield  {author} {\bibinfo {author} {\bibfnamefont {A.}~\bibnamefont
  {de~Visser}}, \bibinfo {author} {\bibfnamefont {F.~E.}\ \bibnamefont
  {Kayzel}}, \bibinfo {author} {\bibfnamefont {A.~A.}\ \bibnamefont
  {Menovsky}}, \bibinfo {author} {\bibfnamefont {J.~J.~M.}\ \bibnamefont
  {Franse}}, \bibinfo {author} {\bibfnamefont {J.}~\bibnamefont {van~den
  Berg}}, \ and\ \bibinfo {author} {\bibfnamefont {G.~J.}\ \bibnamefont
  {Nieuwenhuys}},\ }\href {\doibase 10.1103/PhysRevB.34.8168} {\bibfield
  {journal} {\bibinfo  {journal} {Phys. Rev. B}\ }\textbf {\bibinfo {volume}
  {34}},\ \bibinfo {pages} {8168} (\bibinfo {year} {1986})}\BibitemShut
  {NoStop}%
\bibitem [{\citenamefont {Kuwahara}\ \emph {et~al.}(1997)\citenamefont
  {Kuwahara}, \citenamefont {Amitsuka}, \citenamefont {Sakakibara},
  \citenamefont {Suzuki}, \citenamefont {Nakamura}, \citenamefont {Goto},
  \citenamefont {Mihalik}, \citenamefont {Menovsky}, \citenamefont
  {de~Visser},\ and\ \citenamefont
  {Franse}}]{Kuwahara97_Alpha_elastic_URu2Si2}%
  \BibitemOpen
  \bibfield  {author} {\bibinfo {author} {\bibfnamefont {K.}~\bibnamefont
  {Kuwahara}}, \bibinfo {author} {\bibfnamefont {H.}~\bibnamefont {Amitsuka}},
  \bibinfo {author} {\bibfnamefont {T.}~\bibnamefont {Sakakibara}}, \bibinfo
  {author} {\bibfnamefont {O.}~\bibnamefont {Suzuki}}, \bibinfo {author}
  {\bibfnamefont {S.}~\bibnamefont {Nakamura}}, \bibinfo {author}
  {\bibfnamefont {T.}~\bibnamefont {Goto}}, \bibinfo {author} {\bibfnamefont
  {M.}~\bibnamefont {Mihalik}}, \bibinfo {author} {\bibfnamefont {A.~A.}\
  \bibnamefont {Menovsky}}, \bibinfo {author} {\bibfnamefont {A.}~\bibnamefont
  {de~Visser}}, \ and\ \bibinfo {author} {\bibfnamefont {J.~J.~M.}\
  \bibnamefont {Franse}},\ }\bibfield  {booktitle} {\emph {\bibinfo {booktitle}
  {Journal of the Physical Society of Japan}},\ }\href {\doibase
  10.1143/JPSJ.66.3251} {\bibfield  {journal} {\bibinfo  {journal} {Journal of
  the Physical Society of Japan}\ }\textbf {\bibinfo {volume} {66}},\ \bibinfo
  {pages} {3251} (\bibinfo {year} {1997})}\BibitemShut {NoStop}%
\bibitem [{\citenamefont {McElfresh}\ \emph {et~al.}(1987)\citenamefont
  {McElfresh}, \citenamefont {Thompson}, \citenamefont {Willis}, \citenamefont
  {Maple}, \citenamefont {Kohara},\ and\ \citenamefont
  {Torikachvili}}]{McElfresh87_Pressure_URu2Si2}%
  \BibitemOpen
  \bibfield  {author} {\bibinfo {author} {\bibfnamefont {M.~W.}\ \bibnamefont
  {McElfresh}}, \bibinfo {author} {\bibfnamefont {J.~D.}\ \bibnamefont
  {Thompson}}, \bibinfo {author} {\bibfnamefont {J.~O.}\ \bibnamefont
  {Willis}}, \bibinfo {author} {\bibfnamefont {M.~B.}\ \bibnamefont {Maple}},
  \bibinfo {author} {\bibfnamefont {T.}~\bibnamefont {Kohara}}, \ and\ \bibinfo
  {author} {\bibfnamefont {M.~S.}\ \bibnamefont {Torikachvili}},\ }\href
  {\doibase 10.1103/PhysRevB.35.43} {\bibfield  {journal} {\bibinfo  {journal}
  {Phys. Rev. B}\ }\textbf {\bibinfo {volume} {35}},\ \bibinfo {pages} {43}
  (\bibinfo {year} {1987})}\BibitemShut {NoStop}%
\bibitem [{\citenamefont {{Jeffries}}\ \emph {et~al.}(2010)\citenamefont
  {{Jeffries}}, \citenamefont {{Butch}}, \citenamefont {{Hamlin}},
  \citenamefont {{Sinogeikin}}, \citenamefont {{Evans}},\ and\ \citenamefont
  {{Maple}}}]{Jeffries2010arXiv1002.2245J}%
  \BibitemOpen
  \bibfield  {author} {\bibinfo {author} {\bibfnamefont {J.~R.}\ \bibnamefont
  {{Jeffries}}}, \bibinfo {author} {\bibfnamefont {N.~P.}\ \bibnamefont
  {{Butch}}}, \bibinfo {author} {\bibfnamefont {J.~J.}\ \bibnamefont
  {{Hamlin}}}, \bibinfo {author} {\bibfnamefont {S.~V.}\ \bibnamefont
  {{Sinogeikin}}}, \bibinfo {author} {\bibfnamefont {W.~J.}\ \bibnamefont
  {{Evans}}}, \ and\ \bibinfo {author} {\bibfnamefont {M.~B.}\ \bibnamefont
  {{Maple}}},\ }\href@noop {} {\bibfield  {journal} {\bibinfo  {journal} {arXiv
  e-prints}\ ,\ \bibinfo {eid} {arXiv:1002.2245}} (\bibinfo {year} {2010})},\
  \Eprint {http://arxiv.org/abs/1002.2245} {arXiv:1002.2245 [cond-mat.str-el]}
  \BibitemShut {NoStop}%
\bibitem [{\citenamefont {Shapiro}(2006)}]{ShapiroPhD2016}%
  \BibitemOpen
  \bibfield  {author} {\bibinfo {author} {\bibfnamefont {M.~C.}\ \bibnamefont
  {Shapiro}},\ }\emph {\bibinfo {title} {ELASTORESISTIVITY AS A PROBE OF
  ELECTRONICALLY DRIVEN ROTATIONAL SYMMETRY BREAKING AND ITS APPLICATION TO THE
  HIDDEN ORDER STATE IN URu2Si2}},\ \href@noop {} {Ph.D. thesis},\ \bibinfo
  {school} {Standford University} (\bibinfo {year} {2006})\BibitemShut
  {NoStop}%
\bibitem [{\citenamefont {Choi}\ \emph {et~al.}(2018)\citenamefont {Choi},
  \citenamefont {Ivashko}, \citenamefont {Dennler}, \citenamefont {Aoki},
  \citenamefont {von Arx}, \citenamefont {Gerber}, \citenamefont {Gutowski},
  \citenamefont {Fischer}, \citenamefont {Strempfer}, \citenamefont
  {v.~Zimmermann},\ and\ \citenamefont {Chang}}]{Choi_PRB2018_pressure}%
  \BibitemOpen
  \bibfield  {author} {\bibinfo {author} {\bibfnamefont {J.}~\bibnamefont
  {Choi}}, \bibinfo {author} {\bibfnamefont {O.}~\bibnamefont {Ivashko}},
  \bibinfo {author} {\bibfnamefont {N.}~\bibnamefont {Dennler}}, \bibinfo
  {author} {\bibfnamefont {D.}~\bibnamefont {Aoki}}, \bibinfo {author}
  {\bibfnamefont {K.}~\bibnamefont {von Arx}}, \bibinfo {author} {\bibfnamefont
  {S.}~\bibnamefont {Gerber}}, \bibinfo {author} {\bibfnamefont
  {O.}~\bibnamefont {Gutowski}}, \bibinfo {author} {\bibfnamefont {M.~H.}\
  \bibnamefont {Fischer}}, \bibinfo {author} {\bibfnamefont {J.}~\bibnamefont
  {Strempfer}}, \bibinfo {author} {\bibfnamefont {M.}~\bibnamefont
  {v.~Zimmermann}}, \ and\ \bibinfo {author} {\bibfnamefont {J.}~\bibnamefont
  {Chang}},\ }\href {\doibase 10.1103/PhysRevB.98.241113} {\bibfield  {journal}
  {\bibinfo  {journal} {Phys. Rev. B}\ }\textbf {\bibinfo {volume} {98}},\
  \bibinfo {pages} {241113} (\bibinfo {year} {2018})}\BibitemShut {NoStop}%
\bibitem [{\citenamefont {{Ishida}}\ \emph {et~al.}(2018)\citenamefont
  {{Ishida}}, \citenamefont {{Tsujii}}, \citenamefont {{Hosoi}}, \citenamefont
  {{Mizukami}}, \citenamefont {{Ishida}}, \citenamefont {{Iyo}}, \citenamefont
  {{Eisaki}}, \citenamefont {{Wolf}}, \citenamefont {{Grube}}, \citenamefont
  {{L{\"o}hneysen}}, \citenamefont {{Fernandes}},\ and\ \citenamefont
  {{Shibauchi}}}]{Ishida2018}%
  \BibitemOpen
  \bibfield  {author} {\bibinfo {author} {\bibfnamefont {K.}~\bibnamefont
  {{Ishida}}}, \bibinfo {author} {\bibfnamefont {M.}~\bibnamefont {{Tsujii}}},
  \bibinfo {author} {\bibfnamefont {S.}~\bibnamefont {{Hosoi}}}, \bibinfo
  {author} {\bibfnamefont {Y.}~\bibnamefont {{Mizukami}}}, \bibinfo {author}
  {\bibfnamefont {S.}~\bibnamefont {{Ishida}}}, \bibinfo {author}
  {\bibfnamefont {A.}~\bibnamefont {{Iyo}}}, \bibinfo {author} {\bibfnamefont
  {H.}~\bibnamefont {{Eisaki}}}, \bibinfo {author} {\bibfnamefont
  {T.}~\bibnamefont {{Wolf}}}, \bibinfo {author} {\bibfnamefont
  {K.}~\bibnamefont {{Grube}}}, \bibinfo {author} {\bibfnamefont {H.~v.}\
  \bibnamefont {{L{\"o}hneysen}}}, \bibinfo {author} {\bibfnamefont {R.~M.}\
  \bibnamefont {{Fernandes}}}, \ and\ \bibinfo {author} {\bibfnamefont
  {T.}~\bibnamefont {{Shibauchi}}},\ }\href@noop {} {\bibfield  {journal}
  {\bibinfo  {journal} {arXiv e-prints}\ ,\ \bibinfo {eid} {arXiv:1812.05267}}
  (\bibinfo {year} {2018})},\ \Eprint {http://arxiv.org/abs/1812.05267}
  {arXiv:1812.05267 [cond-mat.supr-con]} \BibitemShut {NoStop}%
\bibitem [{\citenamefont {Hardy}\ \emph {et~al.}(2016)\citenamefont {Hardy},
  \citenamefont {B\"ohmer}, \citenamefont {de' Medici}, \citenamefont {Capone},
  \citenamefont {Giovannetti}, \citenamefont {Eder}, \citenamefont {Wang},
  \citenamefont {He}, \citenamefont {Wolf}, \citenamefont {Schweiss},
  \citenamefont {Heid}, \citenamefont {Herbig}, \citenamefont {Adelmann},
  \citenamefont {Fisher},\ and\ \citenamefont
  {Meingast}}]{Hardy16_K-doped-Pnictide}%
  \BibitemOpen
  \bibfield  {author} {\bibinfo {author} {\bibfnamefont {F.}~\bibnamefont
  {Hardy}}, \bibinfo {author} {\bibfnamefont {A.~E.}\ \bibnamefont {B\"ohmer}},
  \bibinfo {author} {\bibfnamefont {L.}~\bibnamefont {de' Medici}}, \bibinfo
  {author} {\bibfnamefont {M.}~\bibnamefont {Capone}}, \bibinfo {author}
  {\bibfnamefont {G.}~\bibnamefont {Giovannetti}}, \bibinfo {author}
  {\bibfnamefont {R.}~\bibnamefont {Eder}}, \bibinfo {author} {\bibfnamefont
  {L.}~\bibnamefont {Wang}}, \bibinfo {author} {\bibfnamefont {M.}~\bibnamefont
  {He}}, \bibinfo {author} {\bibfnamefont {T.}~\bibnamefont {Wolf}}, \bibinfo
  {author} {\bibfnamefont {P.}~\bibnamefont {Schweiss}}, \bibinfo {author}
  {\bibfnamefont {R.}~\bibnamefont {Heid}}, \bibinfo {author} {\bibfnamefont
  {A.}~\bibnamefont {Herbig}}, \bibinfo {author} {\bibfnamefont
  {P.}~\bibnamefont {Adelmann}}, \bibinfo {author} {\bibfnamefont {R.~A.}\
  \bibnamefont {Fisher}}, \ and\ \bibinfo {author} {\bibfnamefont
  {C.}~\bibnamefont {Meingast}},\ }\href {\doibase 10.1103/PhysRevB.94.205113}
  {\bibfield  {journal} {\bibinfo  {journal} {Phys. Rev. B}\ }\textbf {\bibinfo
  {volume} {94}},\ \bibinfo {pages} {205113} (\bibinfo {year}
  {2016})}\BibitemShut {NoStop}%
\bibitem [{\citenamefont {Boehmer}(2014)}]{AnneThesis2014}%
  \BibitemOpen
  \bibfield  {author} {\bibinfo {author} {\bibfnamefont {A.~E.}\ \bibnamefont
  {Boehmer}},\ }\emph {\bibinfo {title} {\textup{Competing phases in iron-based
  superconductors studied by high-resolution thermal-expansion and
  shear-modulus measurements}}},\ \href
  {http://inis.iaea.org/search/search.aspx?orig_q=RN:47116984; :
  https://publikationen.bibliothek.kit.edu/1000042623/3183263} {Ph.D. thesis},\
  \bibinfo  {school} {Kalrsuhe Institute of Technology} (\bibinfo {year}
  {2014})\BibitemShut {NoStop}%
\bibitem [{\citenamefont {Eilers}\ \emph {et~al.}(2016)\citenamefont {Eilers},
  \citenamefont {Grube}, \citenamefont {Zocco}, \citenamefont {Wolf},
  \citenamefont {Merz}, \citenamefont {Schweiss}, \citenamefont {Heid},
  \citenamefont {Eder}, \citenamefont {Yu}, \citenamefont {Zhu}, \citenamefont
  {Si}, \citenamefont {Shibauchi},\ and\ \citenamefont
  {L\"ohneysen}}]{Eilers16_QO_AFe2As2}%
  \BibitemOpen
  \bibfield  {author} {\bibinfo {author} {\bibfnamefont {F.}~\bibnamefont
  {Eilers}}, \bibinfo {author} {\bibfnamefont {K.}~\bibnamefont {Grube}},
  \bibinfo {author} {\bibfnamefont {D.~A.}\ \bibnamefont {Zocco}}, \bibinfo
  {author} {\bibfnamefont {T.}~\bibnamefont {Wolf}}, \bibinfo {author}
  {\bibfnamefont {M.}~\bibnamefont {Merz}}, \bibinfo {author} {\bibfnamefont
  {P.}~\bibnamefont {Schweiss}}, \bibinfo {author} {\bibfnamefont
  {R.}~\bibnamefont {Heid}}, \bibinfo {author} {\bibfnamefont {R.}~\bibnamefont
  {Eder}}, \bibinfo {author} {\bibfnamefont {R.}~\bibnamefont {Yu}}, \bibinfo
  {author} {\bibfnamefont {J.-X.}\ \bibnamefont {Zhu}}, \bibinfo {author}
  {\bibfnamefont {Q.}~\bibnamefont {Si}}, \bibinfo {author} {\bibfnamefont
  {T.}~\bibnamefont {Shibauchi}}, \ and\ \bibinfo {author} {\bibfnamefont
  {H.~v.}\ \bibnamefont {L\"ohneysen}},\ }\href {\doibase
  10.1103/PhysRevLett.116.237003} {\bibfield  {journal} {\bibinfo  {journal}
  {Phys. Rev. Lett.}\ }\textbf {\bibinfo {volume} {116}},\ \bibinfo {pages}
  {237003} (\bibinfo {year} {2016})}\BibitemShut {NoStop}%
\bibitem [{\citenamefont {{Ishida}}\ \emph {et~al.}(2019)\citenamefont
  {{Ishida}}, \citenamefont {{Hosoi}}, \citenamefont {{Teramoto}},
  \citenamefont {{Usui}}, \citenamefont {{Mizukami}}, \citenamefont {{Itaka}},
  \citenamefont {{Matsuda}}, \citenamefont {{Watanabe}},\ and\ \citenamefont
  {{Shibauchi}}}]{Ishida2019arXiv190807167I}%
  \BibitemOpen
  \bibfield  {author} {\bibinfo {author} {\bibfnamefont {K.}~\bibnamefont
  {{Ishida}}}, \bibinfo {author} {\bibfnamefont {S.}~\bibnamefont {{Hosoi}}},
  \bibinfo {author} {\bibfnamefont {Y.}~\bibnamefont {{Teramoto}}}, \bibinfo
  {author} {\bibfnamefont {T.}~\bibnamefont {{Usui}}}, \bibinfo {author}
  {\bibfnamefont {Y.}~\bibnamefont {{Mizukami}}}, \bibinfo {author}
  {\bibfnamefont {K.}~\bibnamefont {{Itaka}}}, \bibinfo {author} {\bibfnamefont
  {Y.}~\bibnamefont {{Matsuda}}}, \bibinfo {author} {\bibfnamefont
  {T.}~\bibnamefont {{Watanabe}}}, \ and\ \bibinfo {author} {\bibfnamefont
  {T.}~\bibnamefont {{Shibauchi}}},\ }\href@noop {} {\bibfield  {journal}
  {\bibinfo  {journal} {arXiv e-prints}\ ,\ \bibinfo {eid} {arXiv:1908.07167}}
  (\bibinfo {year} {2019})},\ \Eprint {http://arxiv.org/abs/1908.07167}
  {arXiv:1908.07167 [cond-mat.supr-con]} \BibitemShut {NoStop}%
\end{thebibliography}%

\end{document}